\documentclass[prl,preprintnumbers,amsmath,amssymb,superscriptaddress,twocolumn]{revtex4}
\usepackage{graphicx}% Include figure files
\usepackage{dcolumn}% Align table columns on decimal point
\usepackage{bm}% bold math
\usepackage{color}
\usepackage[latin9]{inputenc}
\usepackage{amsmath}
\usepackage{multirow}
\usepackage{color}

\usepackage{siunitx}

\graphicspath{{/}}

\newcommand{\ket}[2][]{\mathinner{\lvert#2\rangle}_{\hspace{-0.1em}#1}}

\newcommand{\mean}[1]{\langle #1 \rangle}
\newcommand{\ayop}{\hat{a}_y}
\newcommand{\adyop}{\hat{a}^\dagger_y}

\begin{document}
\title{Cooling nanofiber-trapped atoms to their quantum ground state}
\title{Preparation of atoms in their motional ground state, \SI{300}{nm} away from a hot surface}
\title{Ground-state cooling of atoms interfaced with a nanophotonic structure}
\title{Ground-state cooling of atoms \SI{300}{nm} away from a hot surface}
\title{Cooling atoms close to their motional ground state \SI{300}{nm} away from a hot surface}
\title{Near-ground-state cooling of atoms optically trapped \SI{300}{nm} away from a hot surface}

\author{Y.~Meng}
\author{A.~Dareau}
\author{P.~Schneeweiss}
\email{philipp.schneeweiss@tuwien.ac.at}
\author{A.~Rauschenbeutel}
\email{arno.rauschenbeutel@tuwien.ac.at}

\affiliation{%
 Vienna Center for Quantum Science and Technology,\\
 TU Wien -- Atominstitut, Stadionallee 2, 1020 Vienna, Austria
}%

\date{\today}
\begin{abstract}
Laser-cooled atoms coupled to nanophotonic structures constitute a powerful research platform for the exploration of new regimes of light--matter interaction. While the initialization of the atomic internal degrees of freedom in these systems has been achieved, a full preparation of the atomic quantum state also requires controlling the center of mass motion of the atoms at the quantum level. Obtaining such control is not straightforward, due to the close vicinity of the atoms to the photonic system that is at ambient temperature. Here, we demonstrate cooling of individual neutral Cesium atoms, that are optically interfaced with light in an optical nanofiber, preparing them close to their three-dimensional motional ground state. The atoms are localized less than \SI{300}{nm} away from the hot fiber surface. Ground-state preparation is achieved by performing degenerate Raman cooling, and the atomic temperature is inferred from the analysis of heterodyne fluorescence spectroscopy signals. Our cooling method can be implemented either with externally applied or guided light fields. Moreover, it relies on polarization gradients which naturally occur for strongly confined guided optical fields. Thus, this method can be implemented in any trap based on nanophotonic structures. Our results provide an ideal starting point for the study of novel effects such as light-induced self-organization, the measurement of novel optical forces, and the investigation of heat transfer at the nanoscale using quantum probes.
\end{abstract}

\maketitle

%, for the precise probing surface effects, and for applications in quantum networking.

%The physical properties of individual laser-cooled atoms such as their resonance wavelength or their response to magnetic fields are precisely defined and can be predicted from first-principle calculations. This makes ideal for applications as quantum sensors 

%%%%%%%%%%%%%%%%%%%%%%%%%%%%%%%%%%%%%%%%%%%%%%%%%%%%%%%%%%%%%%%%%%%%%%%%%%%%%%%%%%%%%%%%%%%%%%%%%%%%%%%%%%%%%%%%%%%%%%%%%%%%%%%%%
%%%  Introduction
%%%%%%%%%%%%%%%%%%%%%%%%%%%%%%%%%%%%%%%%%%%%%%%%%%%%%%%%%%%%%%%%%%%%%%%%%%%%%%%%%%%%%%%%%%%%%%%%%%%%%%%%%%%%%%%%%%%%%%%%%%%%%%%%%

Trapped laser-cooled atoms constitute a versatile experimental platform, with applications ranging from precision measurements~\cite{Cronin09, Ludlow15}, to quantum simulations of solid-state systems~\cite{Bloch12}, and quantum communication~\cite{Kimble08}. A promising pathway is to interface cold-atoms with light confined in nanophotonic structures, which offer a flexible design of the optical modes, very large coupling strengths, and enable the exploration of novel regimes of light--matter interaction~\cite{Lodahl17}. Interfacing atoms with light guided in a nanophotonic element is typically achieved via the evanescent part of the optical mode. As these fields decay on the wavelength-scale, emitters have to be very close to the surface of the photonic structure for purposes of reaching significant coupling strengths.

In order to harness the full potential of cold atoms and to build on their excellent coherence properties, the preparation of the atoms in a well-defined quantum-mechanical state is required. The individual tasks of either trapping cold atoms close to nanophotonic structures~\cite{Vetsch10,Thompson13b} or quantum state preparation~\cite{Bouchoule99,Foerster09} of atoms confined in conventional, free-space traps has already been demonstrated. A quantum-level control of atoms in nanophotonic traps, however, constitutes a prime challenge as various effects such as Johnson noise~\cite{Henkel99}, patch potentials~\cite{McGuirk04}, or coupling to the phononic modes of the structure~\cite{Henkel99b,Wuttke13b}, might lead to decoherence and heating. On a more general note, the very possibility to prepare quantum states including the external degrees of freedom (DOFs) of atoms close to a hot, macroscopic object cannot be taken for granted. In this regard, cooling atoms to the motional ground state constitutes a prerequisite for the preparation of more complex states. So far, only the cooling of a single DOF has been shown close to a nanophotonic structure using microwave or Raman sideband cooling~\cite{Albrecht16,Ostfeldt17}.

Here we demonstrate three-dimensional cooling of individual Cesium atoms that are trapped less than \SI{300}{nm} away from the surface of a solid at ambient temperature, close to their motional ground-state. The latter is the nanofiber-part of a tapered optical fiber. The experiment is carried out using a nanofiber-based optical dipole trap~\cite{Vetsch10}, in which we implement degenerate Raman cooling (DRC)~\cite{Hamann98, Vuletic98, Kerman00, Groebner17, Hu17}, taking advantage of intrinsic trap properties. Cooling can be achieved either using free-space light fields or with fiber-guided light fields only. When cooling is applied continuously, the lifetime of atoms in the trap is increased by one order of magnitude and exceeds \SI{1}{s}, reaching the limit imposed by the background gas pressure in our setup. We use heterodyne fluorescence spectroscopy~\cite{Jessen92} in order to precisely characterize the trapping potential, giving access to the temperature of the atoms and indicating cooling close to the motional ground-state. Our work shows that full control of the three-dimensional motional state of the atoms at the level of single quanta can still be achieved despite the close vicinity to a hot, macroscopic body.

%%%%%%%%%%%%%%%%%%%%%%%%%%%%%%%%%%%%%%%%%%%%%%%%%%%%%%%%%%%%%%%%%%%%%%%%%%%%%%%%%%%%%%%%%%%%%%%%%%%%%%%%%%%%%%%%%%%%%%%%%%%%%%%%%
%%%  Discussion of experimental implementation
%%%%%%%%%%%%%%%%%%%%%%%%%%%%%%%%%%%%%%%%%%%%%%%%%%%%%%%%%%%%%%%%%%%%%%%%%%%%%%%%%%%%%%%%%%%%%%%%%%%%%%%%%%%%%%%%%%%%%%%%%%%%%%%%%

\begin{figure*}
	\includegraphics[width=0.9\textwidth]{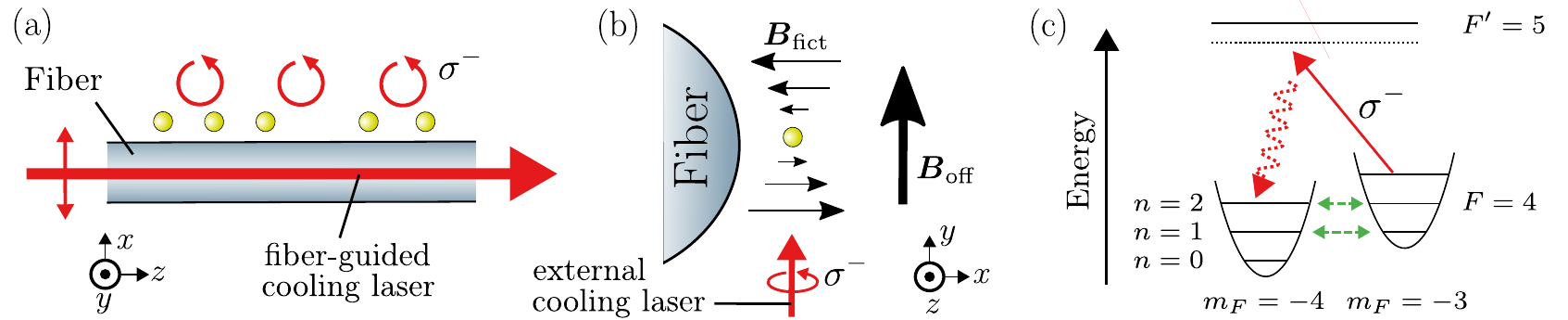}
	\caption{(a) Schematic of laser-cooled Cesium atoms (yellow spheres) interfaced with the evanescent part of a nanofiber-guided cooling light field (red arrows). The atoms are located in the $(x,z)$ plane. The polarization of the cooling light field at the position of the atoms is shown as red curved arrows. It is almost perfectly circular, corresponds to $\sigma^-$ polarization~\cite{Mitsch14a}, and can be used to perform cooling. (b) Illustration of fictitious magnetic field, $\bm{B}_{\rm fict}$, induced by the blue-detuned trapping light field (black horizontal arrows) and homogeneous external offset magnetic field, $\bm{B}_{\rm off}$, (vertical black arrow) for an atom located at the center of a nanofiber-based two-color trapping site. The external cooling laser, propagating along the $y$-axis, is indicated with a red arrow. (c) Schematic of degenerate Raman cooling (DRC). Each Zeeman substate, $\ket{m_F}$, here shown for two states of the $F=4$ hyperfine manifold, has an associated set of motional states, $\ket{n}$, here sketched for a simple 1D harmonic oscillator. The offset magnetic field $\bm{B}_{\rm off}$ can be used to tune different states $\ket{m_F,n}$ and $\ket{m_F',n'}$ ($m_F \neq m_F'$) into resonance. The coupling between these states (green dashed arrows) originates from the fictitious magnetic field $\bm{B}_{\rm fict}$ gradient. To first order, only the state $\ket{m_F=-4,n=0}$ is not coupled by $\bm{B}_{\rm fict}$. An additional laser field (illustrated as a red arrow) can be used to optically pump atoms into this state. Ideally, all atoms are then cooled down to the state $\ket{m_F=-4,n=0}$.}
	\label{fig:Setup_DRC-Scheme}
\end{figure*}

Our experiment relies on the evanescent part of far off-resonant nanofiber-guided light fields to trap Cesium atoms~\cite{Vetsch10}. The trap is realized at the waist of a tapered optical fiber, where the fiber radius is reduced down to \SI{250}{nm}. Trapping is provided by a quasi-linearly polarized~\cite{LeKien04b} blue-detuned running wave with a free-space wavelength of \SI{783}{\nano\meter} and a power of \SI{17.8}{\milli\watt}, and an orthogonally polarized red-detuned standing-wave at \SI{1064}{\nano\meter} wavelength with a total power of \SI{2.88}{\milli\watt}. The atoms are loaded into two diametric one dimensional arrays of trapping sites, where they are confined in all three spatial dimensions. An \textit{ab initio} calculation of the trapping potential yields frequencies of $\{\omega_x,\,\omega_y,\,\omega_z\}/2\pi=\{136,\,83,\,215\}\,\mathrm{kHz}$, i.e., in the Lamb-Dicke regime for atoms close to the ground-state, and a trap minimum located about \SI{280}{\nano\meter} away from the surface of the nanofiber.% where $\omega_i$ is the trap frequency in the i direction. 

Due to the strong transverse confinement of the trapping light fields in the nanofiber, the atoms experience a strongly spatially varying vector ac Stark shift~\cite{Albrecht16}, also known as a fictitious magnetic field~\cite{Cohen-Tannoudji72}. The polarizations of the red-detuned guided light fields generating the standing wave are such that they do not generate a fictitious magnetic field, so only the contribution of the blue-detuned light field remains. Fig.~\ref{fig:Setup_DRC-Scheme}(b) illustrates the resulting fictitious magnetic field $\bm{B}_{\text{fict}}$ near the trap minimum. There it dominantly points along the $x$-axis, and, to first order, its magnitude varies linearly along $y$, i.e.,  $\bm{B}_{\text{fict}} \approx \, b \, y \, \bm{e}_{x}$, with $\bm{e}_{x}$ the unit vector along $x$ and $b=\SI{1.6}{G\per\micro\meter}$ for our trap configuration. The evolution of the spin and $y$ motional DOFs for an atom in the trap is then governed by the following Hamiltonian: 

%Near the minimum trap potential, the nanofiber trap can be approximated as a harmonic potential. A simplified Hamiltonian of nanofiber trapped atom can be written as follows:

\begin{equation}
 \hat{H} = \hbar \omega_y \adyop\ayop + g_F \mu_B \bm{\hat{F}}\cdot(\bm{B}_{\text{off}}+\bm{B}_{\text{fict}}),
 \label{eq:Hamiltonian_1D_initial}
\end{equation}

\noindent where $\bm{\hat{F}}$ is the total angular momentum operator, $g_F$ is the hyperfine Land\'e factor and $\mu_B$ the Bohr magneton. We have assumed the trap to be harmonic and introduced the corresponding annihilation operator $\ayop$. An additional homogeneous offset magnetic field, $\bm{B}_{\text{off}} = B_{\text{off}} \, \bm{e}_{y}$, points along the $y$-axis, see Fig.~\ref{fig:Setup_DRC-Scheme}(b). Using our expression for $\bm{B}_{\text{fict}}$, the Hamiltonian~\eqref{eq:Hamiltonian_1D_initial} can be rewritten:

\begin{equation}
 \hat{H}/\hbar= \omega_y \adyop\ayop + \Delta_\mathrm{off} \hat{F}_y + \Omega \left( \ayop+ \adyop \right) \left( \hat{F}_{+} + \hat{F}_{-} \right),
 \label{eq:H_spin_motion}
\end{equation}

\noindent where we have introduced $\hat{F}_{+}$ ($\hat{F}_{-}$) as the spin raising (lowering) operator for the $\hat{F}_y$ eigenstates ($\hat{F_x} = [\hat{F}_{+} + \hat{F}_{-}]/2$), and used that $\hat{y}=y_0 (\ayop+ \adyop)$. The second term in~\eqref{eq:H_spin_motion} corresponds to the Zeeman shift induced by  $\bm{B}_{\text{off}}$ ($\Delta_\mathrm{off} \propto  B_{\text{off}}$). The last term comes from the fictitious magnetic field gradient, and induces a coherent coupling between spin and motional DOFs, equivalent to a Raman coupling. The spin-motion coupling strength $\Omega$ is proportional to the magnitude of the fictitious magnetic field gradient $b$. In the following, we note $\ket{m_F,n}$ the eigenstates of $\hat{H}$ in the absence of spin-motion coupling ($\Omega = 0$), where $m_F$ is the projection of the hyperfine atomic spin along $y$, i.e., the magnetic quantum number is specified assuming a $y$-quantization axis. The motional state of the atom in the trap is labeled by $n$.

%We use this spin-motion coupling to implement degenerate Raman cooling, as illustrated in Fig.~\ref{fig:Setup_DRC-Scheme}(b).  

Hamiltonian~\eqref{eq:H_spin_motion} enables degenerate Raman cooling, as illustrated in Fig.~\ref{fig:Setup_DRC-Scheme}(c). The offset magnetic field is tuned so that the energies of the states $\ket{m_F=-4,n}$ and $\ket{m_F=-3,n-1}$ are degenerate. These two states are then resonantly coupled by the spin-motion coupling term, which removes (adds) one quantum of motional energy as the atom precesses to a higher (lower) Zeeman state. To obtain cooling, we continuously apply a $\sigma^-$-polarized light field to pump atoms back to the lower Zeeman state. In the Lamb-Dicke regime, the optical pumping preserves the motional state, and the atom is pumped back to  $\ket{m_F=-4,n-1}$. While this process goes on, atoms accumulate in $\ket{m_F=-4,n=0}$, since this state is not resonantly coupled to any other state through spin-motion coupling. In a similar way, higher-order terms in the series expansion of $\bm{B}_{\text{fict}}$ near the trap minimum can enable the cooling of other motional DOFs (see~\cite{Albrecht16} for a more detailed presentation of the fictitious magnetic field profile in our setup) .

%%%%%%%%%%%%%%%%%%%%%%%%%%%%%%%%%%%%%%%%%%%%%%%%%%%%%%%%%%%%%%%%%%%%%%%%%%%%%%%%%%%%%%%%%%%%%%%%%%%%%%%%%%%%%%%%%%%%%%%%%%%%%%%%%
%%%  Lifetime
%%%%%%%%%%%%%%%%%%%%%%%%%%%%%%%%%%%%%%%%%%%%%%%%%%%%%%%%%%%%%%%%%%%%%%%%%%%%%%%%%%%%%%%%%%%%%%%%%%%%%%%%%%%%%%%%%%%%%%%%%%%%%%%%%

\begin{figure}
	\includegraphics[]{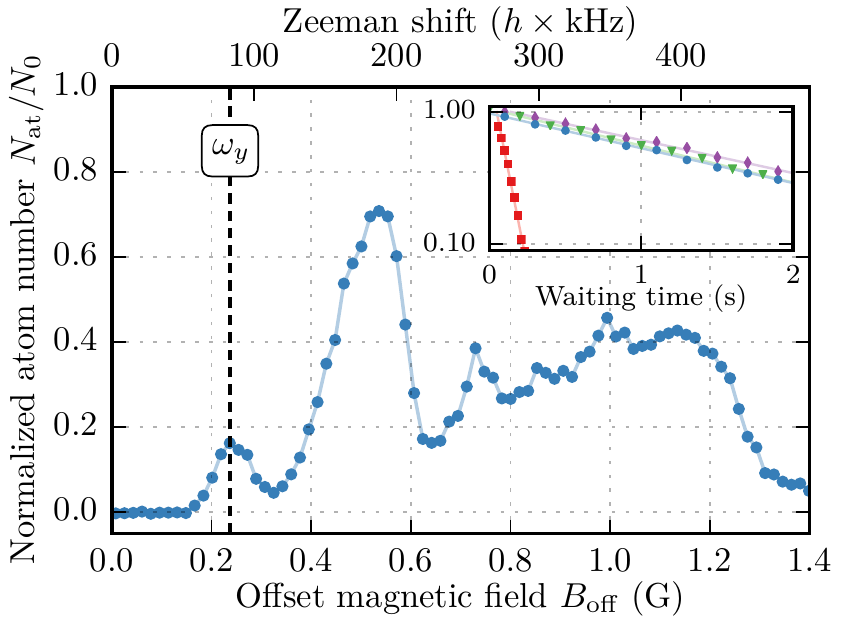}
	\caption{Inset: Normalized number of trapped atoms measured after a variable time, in the absence of cooling (red squares), with DRC cooling using an external (blue circles) or a nanofiber-guided (pink diamonds) laser field, and with polarization gradient cooling (PGC) using the MOT laser fields (green triangles). The measurements with DRC are performed at an offset magnetic field $B_{\rm off}=\SI{0.5}{G}$, with a peak intensity of $I_0=4.1\,I_\mathrm{sat}$ and a detuning of $-12\Gamma$ for the external cooling laser. Here, $I_\mathrm{sat}$ and $\Gamma$ respectively denote the saturation intensity and natural linewidth of the $D_2$ transition of Cesium. The solid lines correspond to fits assuming an exponential decay, yielding a reference lifetime of \SI{75(1)}{ms} in the absence of cooling. In the presence of DRC, we observe lifetimes of \SI{1650(20)}{ms} and \SI{1750(30)}{ms}, respectively with an external and a fiber-guided cooling laser. For PGC, we measure a lifetime of \SI{1560(20)}{ms}. Main graph: Normalized number of atoms still trapped after \SI{500}{ms} of DRC, using an external cooling laser, as a function of $B_{\rm off}$. The corresponding energy shift between two adjacent Zeeman states is given at the top of the figure. Pronounced local maxima of the number of remaining atoms are discernible. We attribute these to resonances in the coupling between motional and spin DOFs, at which the cooling rate is maximized and, thus, the heating of atoms out of the trap efficiently counteracted. A vertical black dashed line indicates the position of the resonance expected from Eq.~\eqref{eq:H_spin_motion}, using our \textit{ab initio} calculation of $\omega_y$.}
	\label{fig:OffsetFieldScan}
\end{figure}

In a typical experimental sequence, atoms are loaded in the nanofiber-based optical trap directly from a magneto-optical trap. A first stage of polarization gradient cooling (PGC) is performed in the trap, and the initial atom number, $N_0$, is inferred from the absorption of a weak nanofiber-guided light field on the cycling transition of the $D_2$ line. The number of atoms remaining in the trap, $N_{\rm at}$, is measured by the same means at the end of the sequence. In order to perform DRC, we apply a $\sigma^-$-polarized laser field, hereafter denoted as the cooling laser. This laser can either propagate in free-space or be guided in the nanofiber. In the former case, we use a laser field propagating along the $+y$-direction and impinging on the atoms from below, see Fig.~\ref{fig:Setup_DRC-Scheme}(b). This external laser field has a $1/e^2$ diameter of \SI{1.4}{mm}, which is enough to cover the full atomic sample. In the latter case, a quasi-linearly polarized fiber-guided laser field is used, whose polarization axis is lying in the plane of the atoms, see Fig.~\ref{fig:Setup_DRC-Scheme}(a). Due to the strong transverse confinement of the light field in the nanofiber, the local polarization is then almost perfectly $\sigma^-$-polarized at the position of the atoms~\cite{Mitsch14a}, thus enabling cooling.

A first signature of cooling is obtained by measuring the lifetime of the atoms in the trap, see inset of Fig.~\ref{fig:OffsetFieldScan}. In this experiment, the fraction of atoms left in the trap $N_{\rm at}/N_0$ is measured after a variable waiting time. Without any cooling (red squares), we observe an exponential decay with a time constant of $\tau_{\rm ref} = \SI{75(1)}{ms}$. When DRC is constantly applied using an external cooling laser (blue circles), the atoms are kept in the trap significantly longer ($\tau_{\rm DRC}^{\rm ext}= \SI{1650(20)}{ms}$), reaching the regime in which collisions with the background gas in the vacuum chamber becomes the limiting factor. This is backed up by an additional measurement (green triangles), in which we constantly apply PGC, which yields a lifetime in reasonable agreement with the DRC value ($\tau_{\rm PGC}= \SI{1560(20)}{ms}$). A similar increase of the lifetime is observed when using a fiber-guided cooling laser to perform DRC (pink diamonds, $\tau_{\rm DRC}^{\rm guided}= \SI{1750(30)}{ms}$). The large improvement of the trap lifetime observed in presence of DRC indicates that this technique can effectively counteract heating mechanisms in the trap. Moreover, it is a first hint that all three motional DOFs are cooled.

\iffalse
tau ref = 75.00 +/- -1.48 ms
tau cool_FG = 1747.57 +/- -31.18 ms
tau MOL = 1560.34 +/- -29.97 ms
tau cool = 1650.62 +/- -19.71 ms
\fi

%A second stage of PGC is applied before the final detection, to avoid any temperature-dependent detection bias [NOTE: this was done for the inset but not for main panel]

We investigate the coupling between spin and motional DOFs by performing DRC for different offset magnetic field strengths, see Fig.~\ref{fig:OffsetFieldScan}, main panel. Specifically, we record the number of atoms in the trap after \SI{500}{ms} of DRC for different values of $B_{\rm off}$. We observe pronounced local maxima, corresponding to the tuning of different motional states of adjacent spin states into resonance. A resonant coupling leads to an increased cooling rate, which results in a longer lifetime, and thus in a larger number of atoms detected. The first local maximum, at $B_{\rm off}\approx \SI{0.25}{G}$, corresponds to a resonant exchange of one spin excitation and one excitation of the azimuthal ($y$) motional DOF, as predicted from the simple model illustrated in Fig.~\ref{fig:Setup_DRC-Scheme}. The expected position of this resonance, deduced from the \textit{ab initio} calculation of $\omega_y$, is indicated by a vertical dashed line in Fig.~\ref{fig:OffsetFieldScan}. Other maxima can be attributed to higher-order couplings terms involving different motional DOFs. 

%For values of $B_{\rm off}$ that lead to a resonance, the DRC rate is maximized, the heating of atoms in the trap is efficiently counteracted, and thus, a large number of atoms remains trapped.

%%%%%%%%%%%%%%%%%%%%%%%%%%%%%%%%%%%%%%%%%%%%%%%%%%%%%%%%%%%%%%%%%%%%%%%%%%%%%%%%%%%%%%%%%%%%%%%%%%%%%%%%%%%%%%%%%%%%%%%%%%%%%%%%%
%%%  Detuning
%%%%%%%%%%%%%%%%%%%%%%%%%%%%%%%%%%%%%%%%%%%%%%%%%%%%%%%%%%%%%%%%%%%%%%%%%%%%%%%%%%%%%%%%%%%%%%%%%%%%%%%%%%%%%%%%%%%%%%%%%%%%%%%%%

\begin{figure}
	\includegraphics[]{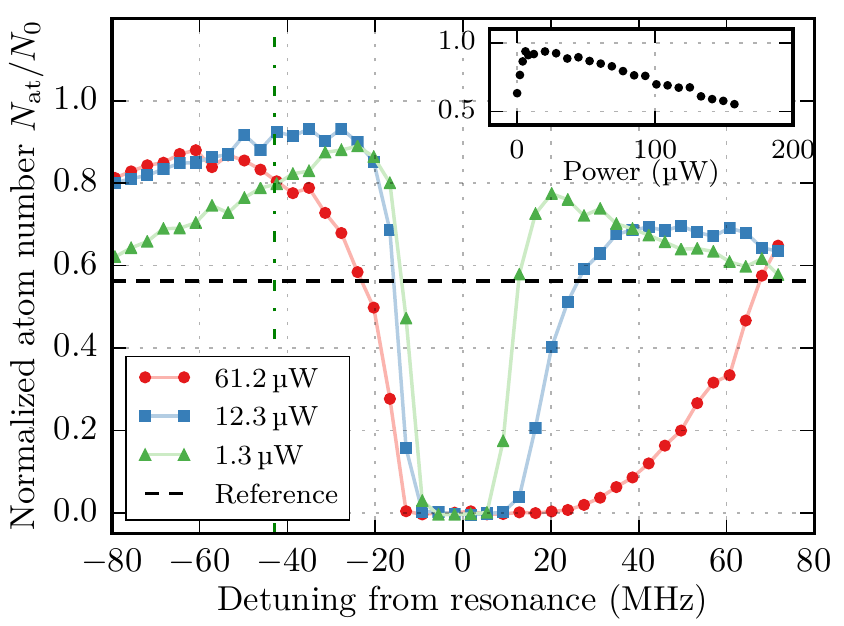}
	\caption{Main graph: Normalized number of atoms remaining in the trap after \SI{80}{ms} of DRC for various detuning of the external cooling laser field. Negative (positive) detuning values correspond to a red (blue) detuning of the laser field with respect to the $(6S_{1/2}, F=4)\to(6P_{3/2}, F'=5)$ optical transition of the trapped Cesium atoms. The measurement is repeated for different power of the external cooling laser, $P_0=$ \SI{61.2}{\micro W} (red circles), \SI{12.3}{\micro W} (blue squares) and \SI{1.3}{\micro W} (green triangles), respectively corresponding to a peak intensity of $I_0/I_\mathrm{sat} = 6.8,\, 1.4\text{ and } 0.14$. The black dashed line indicates the fraction of atoms remaining in the trap in the absence of DRC. When the laser is too close to resonance, the cooling is outweighed by recoil heating, which induces atom losses. Moreover, the signal is asymmetric in the laser detuning, see main text.  Inset: number of remaining atoms in presence of DRC as a function of the cooling laser power. The measurement is taken after \SI{80}{ms} of DRC and at a detuning of $-9.4\,\Gamma$ (vertical line in main graph). The signal is maximized for a power of about \SI{15}{\micro W}, corresponding to $I_0=1.7\,I_\mathrm{sat}$.}
	\label{fig:ProbeScan}
\end{figure}

A careful tuning of the cooling laser's parameters is required to get efficient DRC. In the main panel of Fig.~\ref{fig:ProbeScan}, we show the number of atoms in the trap after \SI{80}{ms} of DRC as a function of the laser detuning, and for three different laser powers. The measurement is performed with an offset magnetic field of \SI{0.5}{G}, and using the external cooling laser. The fraction of atoms left in the absence of DRC is indicated, for reference, with an horizontal dashed black line. A finer scan of the cooling laser power for a fixed detuning is shown in the inset of Fig.~\ref{fig:ProbeScan}. We can clearly identify two limiting regimes. For a low power and a large detuning, the scattering rate is reduced and so is the cooling efficiency, while small detunings and high powers lead to an increased recoil heating rate, which counteracts the cooling and even leads to significant atom losses. 

In addition to changing the scattering rate, scanning the power and detuning modifies the Zeeman-state-dependent ac Stark shift induced by the cooling laser itself, which either increases or reduces the shift induced by the offset magnetic field. The sign of the resulting effective Zeeman shifts depends on the laser detuning, and, close to resonance, their magnitude can be comparable to the energy level spacing in the trap. This can significantly alter the DRC resonance condition and explains the asymmetry of the cooling efficiency for positive and negative detuning in Fig.~\ref{fig:ProbeScan}.

%%%%%%%%%%%%%%%%%%%%%%%%%%%%%%%%%%%%%%%%%%%%%%%%%%%%%%%%%%%%%%%%%%%%%%%%%%%%%%%%%%%%%%%%%%%%%%%%%%%%%%%%%%%%%%%%%%%%%%%%%%%%%%%%%
%%%  Fluorescence spectroscopy
%%%%%%%%%%%%%%%%%%%%%%%%%%%%%%%%%%%%%%%%%%%%%%%%%%%%%%%%%%%%%%%%%%%%%%%%%%%%%%%%%%%%%%%%%%%%%%%%%%%%%%%%%%%%%%%%%%%%%%%%%%%%%%%%%

\begin{figure*}
\includegraphics[width=0.9\textwidth]{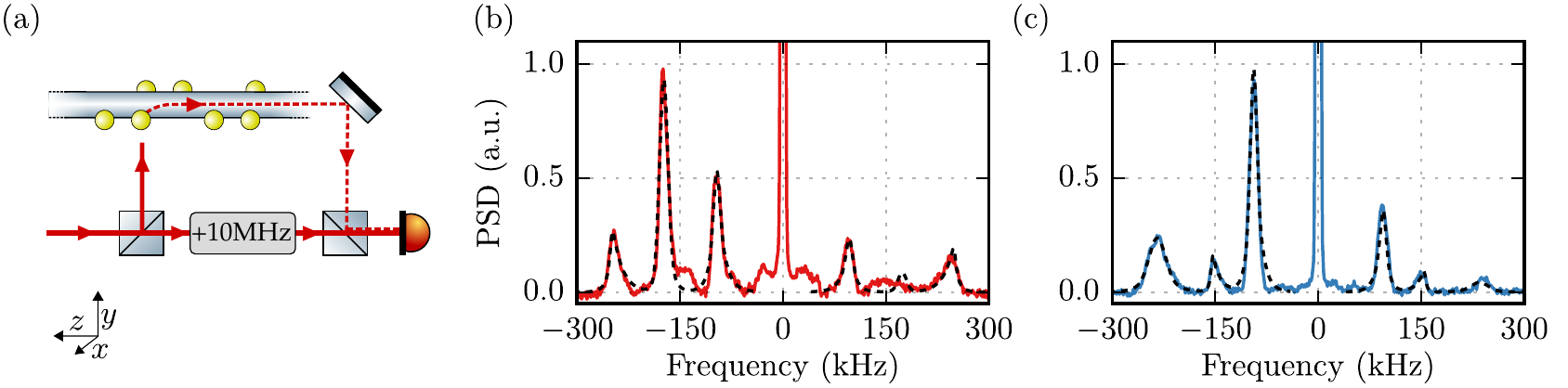}
	\caption{Measurement of trap frequencies and mean numbers of motional quanta of nanofiber-trapped atoms using heterodyne fluorescence spectroscopy (FS). (a) FS setup: The external cooling laser is propagating along the $+y$-direction and impinges on the atoms which are trapped in the $(x,z)$ plane (see also Fig.~\ref{fig:Setup_DRC-Scheme}). During the DRC process, the atoms scatter a fraction of the light into the nanofiber. The light is guided to a beam splitter where it is combined with a reference laser field that is derived from the cooling laser and frequency-shifted by \SI{10}{MHz}. The resulting beatnote is recorded using a SPCM. (b) and (c): Power Spectral Density (PSD) of the SPCM signal for an offset magnetic field of (b) \SI{0.30}{G} and (c) \SI{0.51}{G}. The frequency axis is defined relative to the central beatnote. During a single experimental cycle, the atoms are illuminated for \SI{200}{ms} with an intensity $I \approx 14\,I_\mathrm{sat}$ and a detuning of $-12\,\Gamma$. For each realization, the PSD is obtained from a windowed Fourier transform of the SPCM signal (Welch method, \SI{1}{ms} window). Spectra (b) and (c) are then obtained by averaging the resulting PSDs over around 6000 realizations. The dotted black line corresponds to a fit of the experimental, see main text. From this, we extract mean numbers of excitations of (b) $\{\mean{n_x},\, \mean{n_y},\, \mean{n_z}\} = \{0.10(1),\, 0.78(5),\, 2.5(3)\}$ and (c) $\{\mean{n_x},\, \mean{n_y},\, \mean{n_z}\} = \{1.4(2),\, 0.58(2),\, 0.22(3)\}$, indicating that all motional DOFs can be cooled close to the ground-state.}
	\label{fig:Spectra}
\end{figure*}

More information on the cooling can be obtained by analyzing the light scattered by the atoms. We measure the fluorescence spectrum while performing DRC with the external cooling laser using a heterodyne detection scheme~\cite{Jessen92, Hong06, Das10}. During the DRC process, the atoms scatter light into the nanofiber. This light is guided to a beam splitter where it is combined with a reference laser field, derived from the cooling laser and frequency-shifted by \SI{10}{MHz}, see Fig.~\ref{fig:Spectra}(a). The resulting beatnote is recorded using a SPCM. The motion of the atoms in the trap modulates the scattered light, and gives rise to sidebands at the trap frequencies in the signal's power spectral density (PSD), see Fig.~\ref{fig:Spectra}(b,c). In the weak excitation limit, the scattering process is mostly coherent. The width of the sidebands then depends on the cooling rate, the anharmonicity, and the inhomogeneity of the trapping potential. The temperature of the atomic ensemble can be inferred from the ratio of the amplitudes of the sidebands~\cite{Lindberg86}. For a harmonic oscillator in the Lamb-Dicke regime, the mean number of excitation along $i\in \{x,y,z\}$ is given by $\mean{n_i} = S^{-}_ i / (S^{+}_ i - S^{-}_ i )$, where $S^{\pm}_i$ is the amplitude of the sideband corresponding to the transition $n_i\rightarrow n_i \pm 1$.

% The signal's power spectral density (PSD), as shown on Fig.~\ref{fig:Spectra}(b,c), exhibits a set of three sidebands on both sides of the carrier peak. Indeed, the motion of the atom in the trap leads to a modulation of the scattered light, which results in the appearance of sidebands whose positions correspond to the trap frequencies. % OLD VERSION

In order to derive quantitative information from the recorded spectra, we perform a fit on experimental data. The shape of the sidebands is obtained by calculating the rate $\Gamma_{n_i,n'_i}$ of each individual scattering process $n_i\to n'_i$ using second order perturbation theory~\cite{Javanainen81,Jessen93}. Such a scattering event gives rise to a spectral contribution, whose amplitude and width depend on $\Gamma_{n_i,n'_i}$, and centered around a frequency $f_{n_i,n'_i} = \Delta E_{n_i,n'_i}/h$, where $\Delta E_{n_i,n'_i}$ is the energy difference between the initial and final state, taking into account the trap anharmonicity. All the contributions are then added incoherently to obtain the spectrum. We assume a thermal distribution of the atoms in the trap. We then use the mean number of excitations $\{\mean{n_i}\}_{i=x,y,z}$ and the trap frequencies $\{ \omega_i \}_{i=x,y,z}$ as fit parameters, together with a global scattering rate (setting the minimum width for the sidebands), an offset, and a global amplitude accounting for the combined detection efficiency.

A fit on the spectrum shown in Fig.~\ref{fig:Spectra}(c) yields trap frequencies $\{\omega_x,\, \omega_y,\, \omega_y\}/2\pi = \{154,\, 94,\, 233\}\,\mathrm{kHz}$, which are in reasonable agreement with our \textit{ab initio} calculation. The fitted sideband widths are on the order of ten kilohertz, which sets an upper limit of about \SI{10}{\%} for the inhomogeneity of the trapping frequencies in different sites along the nanofiber. The clear amplitude asymmetry of the Stokes and anti-Stokes sidebands in Fig.~\ref{fig:Spectra}(b,c) is a signature of significant ground-state occupations. For the spectrum shown in Fig.~\ref{fig:Spectra}(b), corresponding to an offset magnetic field of $B_\mathrm{off} = \SI{0.30}{G}$, our fit yields $\{\mean{n_x},\, \mean{n_y},\, \mean{n_z}\} = \{0.10(1),\, 0.78(5),\, 2.5(3)\}$, which corresponds to ground states occupations of \SI{91}{\%} and \SI{56}{\%}, respectively for the radial ($x$) and azimuthal ($y$) motional states. The axial ($z$) motion can be more efficiently cooled by changing the offset magnetic field. A spectrum recorded at $B_\mathrm{off}=\SI{0.51}{G}$ [Fig.~\ref{fig:Spectra}(c)] indicates a mean number of axial excitations of $\mean{n_z} =  0.22(3)$, corresponding to a ground-state occupation of \SI{82}{\%}. For the other motional DOFs, we then find $\{\mean{n_x},\, \mean{n_y}\} = \{1.4(2),\, 0.58(2)\}$.

%%%%%%%%%%%%%%%%%%%%%%%%%%%%%%%%%%%%%%%%%%%%%%%%%%%%%%%%%%%%%%%%%%%%%%%%%%%%%%%%%%%%%%%%%%%%%%%%%%%%%%%%%%%%%%%%%%%%%%%%%%%%%%%%%
%%%  Cooling limits
%%%%%%%%%%%%%%%%%%%%%%%%%%%%%%%%%%%%%%%%%%%%%%%%%%%%%%%%%%%%%%%%%%%%%%%%%%%%%%%%%%%%%%%%%%%%%%%%%%%%%%%%%%%%%%%%%%%%%%%%%%%%%%%%%

We now discuss the mechanisms that can limit the final temperatures reached with our DRC scheme. In the idealized case without heating and neglecting off-resonant excitation, all atoms would end up in the motional ground state after a time that only depends on the cooling rate. With heating, the final temperature, and thus the mean number of motional quanta, is set by a competition between the cooling rate of atoms that have not yet reached the ground state and the rate with which atoms leave the ground state. The cooling rate depends both on the settings of the cooling laser (see Fig.~\ref{fig:ProbeScan}) and on the amplitude of the offset magnetic field (see Fig.~\ref{fig:OffsetFieldScan}). Concerning heating, intrinsic fluctuations of the position and/or steepness of the trap can never be fully avoided in the experiment. This gives rise to a background heating, whose rate was measured to be about $0.3~\mathrm{quanta/ms}$ in the azimuthal ($y$) direction~\cite{Albrecht16}. Moreover, in contrast to many other implementations of DRC, the cooling laser in our experiment is driving an optical cycling transition, $(6S_{1/2},F=4) \to (6P_{3/2},F^\prime=5)$. In this case, the final state in the cooling process, $\ket{m_F=-4, n=0}$, is not a dark state, i.e., it is not decoupled from the laser. We chose this setting since our temperature measurement technique relies on the analysis of light scattered by the atoms. However, this scattering is a source of additional heating due to the transfer of photon-recoil to the atoms. In particular, the intensity of the cooling laser field for the measurements in Fig.~\ref{fig:Spectra} was higher than the optimum intensity indicated in the inset of Fig.~\ref{fig:ProbeScan}. We chose a larger intensity in order to increase the number of fluorescence photons collected and, hence, the signal-to-noise in our spectra. Due to the increased recoil heating in this setting, we are convinced that the measured mean numbers of motional quanta constitute upper bounds of what can be achieved with the DRC method, e.g., when cooling is performed using a light field on a non-cycling transition. We confirmed in additional measurements that, in this case, we obtain comparable boosts in the lifetime of atoms in the trap as shown in the inset in Fig.~\ref{fig:OffsetFieldScan}. 

\iffalse
 - final temp = balance heating & cooling
 - optimum parameters for detecting not the same as cooling : use of cycling transition + too large power ?
 - possibly non optimum setting for B field & non perfect polarization
 - background heating rate then can be limiting
\fi

%%%%%%%%%%%%%%%%%%%%%%%%%%%%%%%%%%%%%%%%%%%%%%%%%%%%%%%%%%%%%%%%%%%%%%%%%%%%%%%%%%%%%%%%%%%%%%%%%%%%%%%%%%%%%%%%%%%%%%%%%%%%%%%%%
%%%  Summary
%%%%%%%%%%%%%%%%%%%%%%%%%%%%%%%%%%%%%%%%%%%%%%%%%%%%%%%%%%%%%%%%%%%%%%%%%%%%%%%%%%%%%%%%%%%%%%%%%%%%%%%%%%%%%%%%%%%%%%%%%%%%%%%%%

In summary, we have shown that degenerate Raman cooling can be efficiently implemented in nanofiber-based optical traps. Remarkably, this technique only requires one additional laser field, which can be fiber-guided, and provides cooling for all three-dimensional motional DOFs. Cooling is enabled by the strong gradients of fictitious magnetic fields, which naturally arise when trapping atoms in evanescent fields~\cite{Albrecht16}. This scheme is thus directly applicable to a vast variety of optical microtraps, and in particular to  traps based on nanophotonic structures. Using a heterodyne fluorescence spectroscopy technique to probe the atomic ensemble temperature, we have confirmed that all motional DOFs can be cooled close to the ground-state, despite the close proximity of the hot fiber surface. This constitutes, to our knowledge, the first experimental evidence of manipulation of all motional DOFs at the quantum level for atoms coupled to a nanophotonic structure. 

Such control is of major importance for cold-atom based nanophotonic devices, where ground state cooling ensures the homogeneity of the atom--waveguide coupling. This could, for instance, improve the performances of atomic Bragg mirrors~\cite{Corzo16,Sorensen16}, quantum memories~\cite{Gouraud15,Sayrin15a}, or squeezing protocols~\cite{Qi16}. The possibility to cool atoms using exclusively guided light fields also opens up interesting opportunities for the design of compact cold-atom based devices, or to cool atoms in cryogenic environments, where optical access is reduced. In a more general context, our results pave the way towards atomic quantum probes for the study of near-surface effects, for example, the experimental study of optical near-field forces~\cite{Scheel15,Sukhov15,Rodriguez-Fortuno15,Kalhor16}, self organization~\cite{Chang13,Griesser13,Eldredge16}, or quantum friction~\cite{Intravaia15}. Ground-state cooling also constitutes a well-defined starting point for the loading of atoms in surface-induced potentials~\cite{Chang14}.

\begin{acknowledgments}
We thank P. Jessen and C. Clausen for stimulating discussions and helpful comments. Financial support by the European Research Council (CoG NanoQuaNt) and the Austrian Science Fund (FWF, SFB NextLite Project No. F 4908-N23 and DK CoQuS project No. W 1210-N16) is gratefully acknowledged.
\end{acknowledgments}

\bibliography{DRC}

\end{document}